# Seeing maximum entropy from the principle of virtual work

Qiuping A. Wang

Institut Supérieur des Matériaux et Mécaniques Avancées du Mans,

44 Av. Bartholdi, 72000 Le Mans, France

**Abstract**

We propose an extension of the principle of virtual work of mechanics to random dynamics of mechanical systems. The total virtual work of the interacting forces and inertial forces on every particle of the system is calculated by considering the motion of each particle. Then according to the principle of Lagrange-d'Alembert for dynamical equilibrium, the vanishing ensemble average of the virtual work gives rise to the thermodynamic equilibrium state with maximization of thermodynamic entropy. This approach establishes a close relationship between the maximum entropy approach for statistical mechanics and a fundamental principle of mechanics, and constitutes an attempt to give the maximum entropy approach, considered by many as only an inference principle based on the subjectivity of probability and entropy, the status of fundamental physics law.

## 1) Introduction

The principle of maximum entropy (maxent) is widely used in the statistical sciences and engineering as a powerful tool and fundamental rule. The maxent approach in statistical mechanics can be traced back to the works of Boltzmann and Gibbs[3] and finally be given the status of principle thanks to the work of Jaynes[4] who used it with Boltzmann-Gibbs-Shannon (BGS) entropy (see below) to derive the canonical probability distribution for statistical mechanics in a simple manner. However, in spite of its success and popularity, maxent has always been at the center of scientific and philosophical discussions and has raised many questions and controversies[4][5][6]. A central question is why a thermodynamic system chooses the equilibrium microstates such that the BGS entropy gets to maximum. As a basic assumption of scientific theory, maxent is not directly or indirectly related to observation and undoubted facts. In the literature, maxent is postulated as such or justified either a priori by the second laws with additional hypothesis such as the entropy functional (Boltzmann or Shannon entropy)[6], or a posteriori by the correctness of the probability distributions derived from it[4]. In statistical inference theory, it was often justified by intuitive arguments based on the subjectivity of probability[4] or by relating it to other principles such as the consistency requirement and the principle of insufficient reason of Laplace, which have been the object of considerable criticisms[5].

Another important question about maxent is whether or not the BGS entropy is unique as the measure of uncertainty or disorder that can be maximized in order to determine probability distributions. This was already an question raised 40 years ago by the scientists who tried to generalize the Shannon entropy by mathematical considerations [9][10].

In the present work, we try to contribute to the debate around maxent by an attempt to derive maxent from a well known fundamental principle of classical mechanics, the virtual work principle or Lagrange-d'Alembert principle (LAP) [1][2] without additional hypotheses to LAP and about entropy property. LAP is widely used in physical sciences as well as in mechanical engineering. It is a basic principle capable of yielding all the basic laws of statics and of dynamics of mechanical systems. It is in addition a simple, clearly defined, easily understandable and palpable law of physics. It is hoped that this derivation is scientifically and pedagogically beneficial for the understanding of maxent and of the relevant questions and controversies around it. In this work, the term entropy, denoted by $S$, is used in the sense of the second law of thermodynamics for equilibrium system.

## 2) Principle of virtual work

The variational calculus in mechanics has a long history which may be traced back to Galilei and other physicists of his time who studied the equilibrium problem of statics with LAP (or virtual displacement[1]). LAP gets unified and concise mathematical forms thanks to Lagrange[1] and d'Alembert[2] and is considered as a most basic principle of mechanics from which all the fundamental laws of statics and dynamics can be understood thoroughly.

LAP says that the total work done by all forces acting on a system in static equilibrium is zero on all possible virtual displacements which are consistent with the constraints of the system. Let us suppose a simple case of a system of $N$ points of mass in equilibrium under the action of $N$ forces $F_i$ ($i=1,2,…N$) with $F_i$ on the point $i$, and imagine virtual displacement of each point $\delta \vec{r}_i$ for the point $i$. According to LAP, the virtual work $\delta W$ of all the forces $F_i$ on all $\delta \vec{r}_i$ cancels itself for static equilibrium, i.e.

$$\delta W = \sum_{i=1}^{N} \vec{F}_i \cdot \delta \vec{r}_i = 0 \quad (1)$$

This principle for statics was extended to dynamical equilibrium by d'Alembert[2] in the LAP by adding the initial force $-m_i \vec{a}_i$ on each point:

$$\delta W = \sum_{i=1}^{N} (\vec{F}_i - m_i \vec{a}_i) \cdot \delta \vec{r}_i = 0 \quad (2)$$

where $m_i$ is the mass of the poin $i$ and $\vec{a}_i$ its acceleration. From this principle, we can not only derive Newtonian equation of dynamics, but also other fundamental principles such as least action principle.

## 3) Why maximum thermodynamic entropy ?

We suppose that the mechanics laws are usable not only for mechanical system containing small number of particles in regular motion, but also for large number of particles in random and stochastic motion for which one has to use statistical approach introducing probability distribution of mechanical states. Let us first consider an ensemble of *equilibrium* systems, each composed of *N* particles in random motion with $\vec{v}_i$ the velocity of the particle *i*. It will be shown that the result for canonical ensemble can be easily extended to microcanonical ensemble and grand-canonical ensemble. Without loss of generality, let us look at a system without macroscopic motion, i.e., $\sum_{i=1}^{N} \vec{v}_i = 0$.

---

[1] In mechanics, the virtual displacement of a system is a kind of imaginary infinitesimal displacement with no time passage and no influence on the forces. It should be perpendicular to the constraint forces.

We imagine that the system in thermodynamic equilibrium leaves the equilibrium state by a reversible infinitesimal virtual process. Let $\vec{F}_i$ be the force on a particle *i* of the system at that moment. $\vec{F}_i$ includes all the interacting forces particles-particles and particles-walls of the container. During the virtual process, each particle with acceleration $\ddot{\vec{r}}_i$ has a virtual displacement $\delta \vec{r}_i$. The total virtual work on this displacement is given by

$$\delta W = \sum_{i=1}^{N} (\vec{F}_i - m\ddot{\vec{r}}_i) \cdot \delta \vec{r}_i \tag{3}$$

Although the sum of the accelerations of all the particles vanishes, i.e., $\sum_{i=1}^{N} m\ddot{\vec{r}}_i = 0$, the acceleration $\ddot{\vec{r}}_i$ on each particle can be nonzero. So in general $\sum_{i=1}^{N} m\ddot{\vec{r}}_i \cdot \delta \vec{r}_i \neq 0$. As a matter of fact, we have $m\ddot{\vec{r}}_i \cdot \delta \vec{r}_i = m\delta \dot{\vec{r}}_i \cdot \dot{\vec{r}}_i = \delta(\frac{1}{2} m \dot{\vec{r}}_i^2) = \delta e_{ki}$ where $e_{ki}$ is the kinetic energy of the particle. On the other hand, we suppose these are no dissipative forces in the system or on the particles. It means that the energy of the system will not change if the system is completely closed and isolated. Let $e_{pi}$ be the potential energy of a particle *i* subject to the force $\vec{F}_i$, we should have $\vec{F}_i = -\nabla e_{pi}$ and

$$\sum_{i=1}^{N} \vec{F}_i \cdot \delta \vec{r}_i = \sum_{i=1}^{N} -\nabla_i e_i \cdot \delta \vec{r}_i = -\sum_{i=1}^{N} \delta e_{pi} \tag{4}$$

So finally it follows that

$$\delta W = -\sum_{i=1}^{N} (\delta e_{pi} + \delta e_{ki}) = -\sum_{i=1}^{N} \delta e_i \tag{5}$$

where $\delta e_i$ is a virtual variation of the total energy $e_i = e_{pi} + e_{ki}$ of the particle *i* and $E = \sum_{i=1}^{N} e_i$ is the total energy of the *N* particles.

At this stage, no statistics has been done. The particles are treated as if they had regular dynamics. As a matter of fact, when the dynamics is random such as in a thermodynamic system, a microscopic process can leads the $N$ particles from a given microstate to different microstates $j$ with different probability $p_j$ ($j=1,2 \ldots w$). If we looks at the system in the phase space, the considered process with given virtual displacements can take different directions or paths each leading to a given microstate with some likelihood. Hence the virtual work given by Eq.(5) is not a correct and complete expression for the random dynamics. It is in fact the virtual work of a possible process leading to a microstate $j$. It should be written as $\delta W_j = -\sum_{i=1}^{N}(\delta e_i)_j$ instead of Eq.(5). This "partial" virtual work cannot be used in the LAP since it is only a possible part of the total virtual work whose correct expression needs the introduction of the probability distribution of microstates $p_j$. Logically, the total virtual work should be an average of the work given by Eq.(5) over all the possible microstates, i.e.,

$$\delta W = \sum_{j=1}^{w} p_j \delta W_j. \tag{6}$$

This expression is essential in the application of LAP, an approach originally for regular dynamics, to irregular and random dynamics. Eq.(6) makes it possible to introduce the dynamic uncertainty (entropy) into the variational approach as shown below. In terms of thermodynamic ensemble, Eq.(6) is the ensemble average of the virtual works of all the members of an ensemble of systems distributed over the microstates. It is this average which is measurable and has a physical sense in the case of random dynamics just as the usual average energy in thermodynamics. It is not conceivable to let the partial virtual work of Eq.(5) vanish because this would signifies that the random motion in each direction in phase space of the virtual process is regular according to LAP and there would be only one direction or phase path of the virtual process leading to only one microstate, which is contradictory with the hypothesis of the random dynamics. This reasoning is the essential difference of the present approach from the simple search for mechanics principle and a direct use of the latter to each possible state or trajectory in phase space. The use of mechanics principle in regular way in general yields regular mechanical laws irrelevant to thermodynamics. The dynamical randomness is the fact that not all the possible states or trajectories follow the regular mechanical laws due to noises, certain chaos, or to quantum mechanics in which the Newtonian laws are obeyed only statistically. The statistical Newtonian second law given in [11] is an example.

Eq.(6) can be accounted for in an explicit way as follows. A microstate $j$ is some distribution of the $N$ particles over the one particle states $k$ with energy $\varepsilon_k$ where $k$ varies from, say, 1 to $g$ ($g$ can be very large). We imagine $N_j$ identical particles distributed over the $g$ states at a microstate $j$ which is here a combination of $g$ numbers $n_k$ of particles over the $g$ states, i.e., $j=\{n_1, n_2, \ldots n_g, \}$. We naturally have $N_j = \sum_{k=1}^{g}(n_k)_j$ and $E_j = \sum_{k=1}^{g}(n_k)_j \varepsilon_k$. During the process of virtual work, only the energy of the particle can change (the fact that virtual work does not affect $n_k$ can be understood from quantum point of view since $\varepsilon_k$ is discrete but virtual work is infinitesimal and continuous). For a given $j$ with probability $p_j$, the virtual work given in Eq.(5) is now

$$\delta W_j = -\sum_k (n_k)_j \delta\varepsilon_k = -\delta \sum_k (n_k)_j \varepsilon_k + \sum_k (\delta n_k)_j \varepsilon_k = -\delta E_j + \sum_k (\delta n_k)_j \varepsilon_k . \qquad (7)$$

The first term of the right hand side is the total energy variation due to the one particle energy variation $\delta\varepsilon_k$ caused by the virtual work as well as to the variation in particle number $\delta N_j$ of the system. The second term is just the energy variation caused by the particle number variation $\delta N_j = \sum_k (\delta n_k)_j$. Hence Eq.(6) reads

$$\delta W = -\sum_j p_j \delta E_j + \sum_j p_j \sum_k \varepsilon_k (\delta n_k)_j = -\overline{\delta E} + \sum_k \varepsilon_k \overline{\delta n_k} = -\overline{\delta E} + \mu \overline{\delta N} . \qquad (8)$$

where we put an expression for the chemical potential $\mu = \sum_k \varepsilon_k \overline{\delta n_k} / \overline{\delta N}$ with $\overline{\delta N} = \sum_j p_j \delta N_j = \sum_j p_j \sum_k (\delta n_k)_j = \sum_k \overline{\delta n_k}$ and $\overline{\delta E} = \sum_j p_j \delta E_j$. Since $\overline{\delta E} = \delta \overline{E} - \sum_j E_j \delta p_j$ and $\overline{\delta N} = \delta \overline{N} - \sum_j N_j \delta p_j$ with $\overline{E} = \sum_j p_j E_j$ and $\overline{N} = \sum_j p_j N_j$, we get

$$\delta W = -\delta \overline{E} + \sum_j E_j \delta p_j + \mu \delta \overline{N} - \mu \sum_j N_j \delta p_j = -\delta \overline{E} + \mu \delta \overline{N} + \sum_j (E_j - \mu N_j) \delta p_j \qquad (9)$$

Now using the first law $\delta \overline{E} = \delta Q - \delta W + \mu \delta \overline{N}$ for Grand-canonical ensemble, we identify the heat transfer $\delta Q = \sum_j (E_j - \mu N_j) \delta p_j$. For a reversible virtual process, we can write $\delta S = \beta \delta Q = \beta \sum_j (E_j - \mu N_j) \delta p_j$ and get

$$\delta W = -\delta \overline{E} + \mu \delta \overline{N} + \frac{\delta S}{\beta}. \qquad (10)$$

where $S$ is the thermodynamic entropy of the second law.

The following variational calculus for different ensemble is straightforward. According to LAP $\delta W = 0$, we have

$$\delta(S - \beta\overline{E} + \beta\mu\overline{N}) = 0 \qquad (11)$$

which is the usual algorithm of maxent for grand-canonical ensemble. The only difference is that here the "constraints" associated with energy and particle number appear in the variational calculus as a simple consequence of LAP, in contrast to the introduction of these constraints in the inference theory or inferential statistical mechanics[4] by the argument that an averaged value of an observable quantity represents a factual information to be put into the maximization of information in order to derive unbiased probability distribution[5].

In order to see further details about this maxent, let us suppose the entropy is a function of the probability distribution $p_j$ of the considered moment, i.e., $S = f(p_1, p_2,...p_j...)$. We can write $\delta S = \sum_j \frac{\partial f}{\partial p_j} \delta p_j$ due to the variations of the virtual process. On the other hand, we have $\delta S = \beta \sum_j (E_j - \mu N_j)\delta p_j$ which implies

$$\sum_j (\frac{\partial f}{\partial p_j} - \beta E_j + \beta\mu N_j)\delta p_j = 0. \qquad (12)$$

By virtue of the normalization condition $\sum_j \delta p_j = 0$, one can prove [12] that

$$\frac{\partial f}{\partial p_j} - \beta E_j + \beta\mu N_j = \alpha. \qquad (13)$$

with a constant $\alpha$. Eq.(13) can be used for deriving the probability distribution of the nonequilibrium component of the dynamics if the functional $f$ is given. Inversely, if the probability distribution is known, one can derive the functional of $S$.

For canonical ensemble, we have $\delta \overline{N} = 0$ and

$$\delta(S - \beta\overline{E}) = 0 \tag{14}$$

or, by the same argument as above,

$$\frac{\partial f}{\partial p_j} - \beta E_j = \alpha. \tag{15}$$

For microcanonical ensemble, the system is completely closed and isolated with constant energy $\delta\overline{E} = 0$ and constant particle number $\delta\overline{N} = 0$. When the virtual displacements occur, the total virtual work would be transformed into virtual heat such that Eq.(10) becomes $\delta W - \delta Q = 0$. LAP $\delta W = 0$ leads to

$$\delta S = 0 \text{ or } \frac{\partial f}{\partial p_j} = \alpha \tag{16}$$

which necessarily yields uniform probability distribution over the different microstates $j$, i.e., $p_j = 1/w$ whatever is the form of the entropy $S$. Note that here the uniform distribution over the microstates is not an a priori assumption, but a consequence of LAP.

This equiprobability can be proven as follows only by supposing that $S = f(p_1, p_2, \ldots p_j \ldots)$ is a strictly increasing or decreasing function of all $p_j$ throughout the interval $0 \leq p_j \leq 1$, i.e., its derivatives $\frac{\partial f}{\partial p_j} > 0 \text{ or } < 0$ and are zero only at some finite number of points on the interval. However, Eq.(16) tells us that $\frac{\partial f}{\partial p_j} = \alpha$ is a constant independent of $p_j$, implying that $S = f(p_1, p_2, \ldots p_j \ldots)$ is either a linear function of all $p_j$, or all $p_j$ are identical. It is evident that entropy cannot be linear function of $p_j$. The equal probability of all microstates follows.

The conclusion of this section is that, at thermodynamic equilibrium, the maxent under the constraints of energy is a consequence of the equilibrium condition LAP extended to random motion. From Eq.(8), one notices that maxent can be written in the following concise form for any ensemble with $n$ random variables $X_i$ ($i=1,2 \ldots n$):

$$\sum_{i=1}^{n} \chi_i \overline{\delta X_i} = 0 \tag{17}$$

where $\chi_i$ is some constant corresponding to $\chi_i$. For grand-canonical ensemble,

this is $\overline{\delta E} - \mu\overline{\delta N} = 0$ and for canonical ensemble, it is $\overline{\delta E} = 0$.

We stress that in the above derivation, the only essential assumptions or fundamental physical hypotheses used before the LAP are the first and second laws of thermodynamics for equilibrium system and reversible process. Hence the three algorithms of maxent for the three statistical ensembles are in principle valid for all systems for which the first and second laws are valid. We would like to mention here that this derivation of maxent is not associated with any given form of entropy like in the original version of Jaynes principle.

### 4) Concluding remarks

This work shows that the maximum entropy principle has a close connection with the fundamental principle of classical mechanics, the principle of virtual work, i.e., for a mechanical system to be in thermodynamics equilibrium with maximum entropy, the total virtual work of all the forces on all the elements (particles) of the system should vanish. Indeed, if one admits that thermodynamic entropy is a measure of dynamical disorder and randomness, it is natural to say that this disorder must get to maximum in order that all the random forces act on each degree of freedom of the motion in such a way that over any possible (virtual) displacement, the work of all the forces is zero. In other words, this vanishing work can be obtained if and only if the randomness of the forces is at maximum over all degree of freedom allowed by the constraints to get stable equilibrium state.

To our opinion, the present result is helpful not only for the understanding of maxent derived from a more basic and well understood mechanical principle, it also shows that entropy in physics is not necessarily a subjective quantity reaching maximum for correct inference, and that maximum entropy is a law of physics but not merely an inference principle.

After finishing this paper, the author became aware of a work of Plastino and Curado[12] on the equivalence between the particular thermodynamic relation $\delta S = \beta\overline{\delta E}$ and maxent in the derivation of probability distribution. They consider the particular thermodynamic process affecting only the microstate population in order to find a different way from maxent to derive probability. The work part is not considered in their work. Their analysis is pertinent and consequential. The present work provides a substantial support of their reasoning from a basic principle of mechanics.